**Some Useful Relations for Analyzing Nanoscale MOSFETs Operating in the Linear Region**

Mark Lundstrom and Xingshu Sun
Purdue University
March 9, 2016

**Abstract** – Several equations used to model and characterize the linear region *IV* characteristics of nanoscale field-effect-transistors are derived. The meaning of carrier mobility at the nanoscale is discussed by defining two related quantities, the apparent mobility and the ballistic mobility. The validity of Matthiessen's Rule for relating the apparent mobility to the ballistic and diffusive mobilities is examined. Other questions that arise in the analysis and characterization of nanoscale field-effect transistors are also discussed. These notes are intended pull together in one place some key equations needed to analyze the linear region performance of nanoscale MOSFETs and to point out errors in some previous publications.

**1. Introduction**
Modern silicon field-effect transistors operate in the quasi-ballistic transport regime, and III-V MOSFETs operate very close to the ballistic limit. The essential physics of well-designed nanoscale transistors is generally understood [1, 2]. These concepts have been embodied in the MIT Virtual Source (VS) model, which has been widely used for technology benchmarking [3]. When analyzing the performance of modern nanoscale transistors, we are confronted with several questions. First, given the material parameters, how does one compute the mobility for a semiconductor? Second, given the measured mobility, how does one deduce the mean-free-path for backscattering? Third, how are familiar concepts like mobility generalized to the nanoscale by introducing an apparent mobility and a ballistic mobility? Another question concerns the validity of Matthiessen's rule when used for combining the scattering limited and ballistic mobilities. These questions and others are addressed in this paper for a 2D semiconductor with parabolic energy bands, but the concepts here are readily extended to more complex bandstructures and to 1D (gate all round) MOSFETs.

This paper compiles in one place several key equations needed to analyze the IV characteristics of nanoscale MOSFETs. For a discussion of the physical underpinnings of the equations, see Lundstrom and Jeong [4]. The notation here follows that of [4]. As summarized at the beginning of Sec. 8, some errors in previous publications are also pointed out.

## 2. The 2D Conductance from the Landauer Perspective

Begin with the Landauer expression for current ([4], eqn. (2.46))

$$I = \frac{2q}{h}\int \mathcal{T}(E) M(E)(f_1 - f_2) dE . \tag{1}$$

Under low bias, the current is ([4], eqn. (2.50))

$$I = \left\{\frac{2q^2}{h}\int \mathcal{T}(E) M(E)\left(-\frac{\partial f_0}{\partial E}\right) dE\right\} V = GV , \tag{2}$$

from which we find the conductance as ([4], eqn. (2.51))

$$G = \frac{I}{V} = \frac{2q^2}{h}\int \mathcal{T}(E) M(E)\left(-\frac{\partial f_0}{\partial E}\right) dE . \tag{3}$$

For diffusive transport, we write the 2D conductance in terms of the 2D conductivity ([4], eqn. (3.3))

$$G_{2D} = \sigma_S \frac{W}{L} . \tag{4}$$

By equating (3) and (4), we get an expression for the **2D conductivity**,

$$\sigma_S = \left\{\frac{2q^2}{h}\int \{\mathcal{T}(E)L\}\{M(E)/W\}\left(-\frac{\partial f_0}{\partial E}\right) dE\right\} . \tag{5}$$

Equation (3) applies from the ballistic to diffusive limits, i.e. from $\mathcal{T}(E)=1$ to $\mathcal{T}(E)\ll 1$, so eqn. (5) also applies from the ballistic to diffusive limits. We should probably call $\sigma_S$ an **apparent conductivity** because conductivity is generally understood to be a concept that applies only in the diffusive limit and should not depend on the sample size.

Next, let's make two definitions so that our notation is consistent with Lundstrom and Jeong [4]. In 2D, the number of channels per unit width is $M(E)/W$, which we will write as ([4], eqn. (2.25))

$$M_{2D}(E) \equiv M(E)/W . \tag{6}$$

The quantity, $\mathcal{T}(E)L$, has the units of length, let's call it the **apparent mean-free-path**,

$$\lambda_{app}(E) \equiv \mathcal{T}(E)L. \tag{7}$$

In the ballistic limit, $\mathcal{T}(E) = 1$, so

$$\lambda_{app}(E) = L \qquad \text{(ballistic)} \tag{8}$$

while in the diffusive limit, $\mathcal{T}(E) = \lambda(E)/L$, so

$$\lambda_{app}(E) = \lambda(E) \qquad \text{(diffusive)}, \tag{9}$$

where $\lambda(E)$ is the mean-free-path for backscattering, MFP, ([4], eqn. (6.16)). In general, the **transmission** is ([4], eqn. (2.43))

$$\mathcal{T}(E) = \frac{\lambda(E)}{\lambda(E) + L}, \tag{10}$$

so from eqn. (7) we find ([4], eqn. (3.40))

$$\lambda_{app}(E) = \mathcal{T}(E)L = \frac{\lambda(E)L}{\lambda(E) + L} = \left(\frac{1}{L} + \frac{1}{\lambda(E)}\right)^{-1}. \tag{11}$$

The apparent MFP is essentially the smaller of two quantities, the MFP for backscattering and the channel length. With these definitions, eqn. (5), the apparent conductivity in 2D, becomes

$$\sigma_{app} = \left\{ \frac{2q^2}{h} \int \lambda_{app}(E) M_{2D}(E) \left( -\frac{\partial f_0}{\partial E} \right) dE \right\}. \tag{12a}$$

Equation (12a) is a clear prescription for computing the apparent conductivity from the known material parameters and sample length. The bandstructure determines $M_{2D}(E)$, the bandstructure and scattering physics determine $\lambda(E)$, and the bandstructure, scattering physics, and channel length determine $\lambda_{app}(E)$. For diffusive sample, $\lambda_{app} \to \lambda$ and (12a) becomes

$$\sigma_s = \left\{ \frac{2q^2}{h} \int \lambda(E) M_{2D}(E) \left( -\frac{\partial f_0}{\partial E} \right) dE \right\}, \tag{12b}$$

which is the standard expression for conductivity but expressed in Landauer form.

## 3. Mobility from Conductivity

Now let's continue and see how to compute the mobility. In a long, diffusive sample, we would write

$$\sigma_S \equiv n_S q \mu_n. \tag{13}$$

We should understand eqn. (13) to be the definition of mobility. More generally, we can define an **apparent mobility**, which is related to the apparent conductivity by

$$\sigma_{app} \equiv n_S q \mu_{app}, \tag{14}$$

By equating eqns. (14) and (12a), we find the apparent mobility as ([4], eqn. (3.35))

$$\mu_{app} \equiv \left(\frac{2q/h}{n_S}\right) \int \lambda_{app}(E) M_{2D}(E) \left(-\frac{\partial f_0}{\partial E}\right) dE. \tag{15}$$

Equation (15) defines the apparent mobility. In the diffusive limit, $\lambda_{app}(E) = \lambda(E)$, where $\lambda(E)$ is the mean-free-path for backscattering, and the scattering limited mobility becomes ([4], eqn. (3.38))

$$\boxed{\mu_n \equiv \left(\frac{2q/h}{n_S}\right) \int \lambda(E) M_{2D}(E) \left(-\frac{\partial f_0}{\partial E}\right) dE}, \tag{16a}$$

which is the **Kubo-Greenwood formula** expressed in Landauer form. In the ballistic limit, $\lambda_{app}(E) = L$, and the **ballistic mobility** becomes ([4], eqn. (3.37))

$$\boxed{\mu_B \equiv \left(\frac{2q/h}{n_S}\right) \int L M_{2D}(E) \left(-\frac{\partial f_0}{\partial E}\right) dE}. \tag{16b}$$

In general, we need to use eqn. (11) for the apparent mean-free-path in eqn. (15), and we find the apparent mobility as

$$\boxed{\mu_{app}(L) \equiv \left(\frac{2q/h}{n_S}\right) \int \left(\frac{1}{L} + \frac{1}{\lambda(E)}\right)^{-1} M_{2D}(E) \left(-\frac{\partial f_0}{\partial E}\right) dE}. \tag{17}$$

## 4. Mean-Free-Path for Backscattering from Mobility

It is often "easy" to measure the mobility, so the question of how we can deduce the MFP for backscattering from the measured mobility arises. Equation (16a), the Kubo-Greenwood formula for mobility expressed in a Landauer form, can also be expressed as

$$\mu_n \equiv \left(\frac{2q/h}{n_S}\right)\langle\langle\lambda(E)\rangle\rangle\langle M_{2D}(E)\rangle, \tag{18}$$

where $\langle\langle\lambda(E)\rangle\rangle$ is the average mean-free-path for backscattering, as defined by eqn. (3.30) of [4] as

$$\langle\langle\lambda(E)\rangle\rangle = \frac{\int \lambda(E)M_{2D}(E)\left(-\frac{\partial f_0}{\partial E}\right)dE}{\int M_{2D}(E)\left(-\frac{\partial f_0}{\partial E}\right)dE}. \tag{19}$$

The quantity,

$$\langle M(E)\rangle = \int M(E)\left(-\frac{\partial f_0}{\partial E}\right)dE, \tag{20}$$

is interpreted as the number of channels in the Fermi window ([4], eqn. (5.3)). To evaluate eqn. (20), recall that for parabolic energy bands the number of channels per unit width is ([4], eqn. (2.31))

$$M_{2D}(E) = g_v \frac{\sqrt{2m^*(E-E_C)}}{\pi\hbar} \text{ m}^{-1}. \tag{21}$$

Using eqn. (21) in (20) we find

$$\langle M_{2D}\rangle = \frac{\sqrt{\pi}}{2} g_v \frac{\sqrt{2m^* k_B T}}{\pi\hbar} \mathcal{F}_{-1/2}(\eta_F), \tag{22}$$

where

$$\eta_F = \frac{E_F - E_C}{k_B T}. \tag{23}$$

Equation (22) is eqn. (3.18) of [4]. For a review of Fermi-Dirac integrals and a discussion of the difference between the Roman $F$ Fermi-Dirac integral and the script $\mathcal{F}$ Fermi-Dirac integral, see [5].

For a single parabolic energy band, the sheet carrier density is ([4], eqn. (3.19))

$$\boxed{n_S = g_v \frac{m^* k_B T}{\pi\hbar^2} \mathcal{F}_0(\eta_F) \text{ m}^{-2}}. \tag{24}$$

Now we can use eqns. (22) and (24) in eqn. (18) to find

$$\boxed{\mu_n = \frac{\upsilon_T \langle\langle \lambda(E) \rangle\rangle}{2(k_B T/q)} \frac{\mathcal{F}_{-1/2}(\eta)}{\mathcal{F}_0(\eta)}}, \quad (25a)$$

Note that eqn. (25a) makes no assumption about the specific form of the energy-dependent mean-free-path for backscattering. It does, however, assume a single parabolic energy band in eqn. (21) and (24).

Given a measured value of mobility, eqn. (25a) permits us to extract the energy-averaged mean-free-path for backscattering from

$$\boxed{\langle\langle \lambda(E) \rangle\rangle = \frac{(2k_B T/q)}{\upsilon_T} \frac{\mathcal{F}_0(\eta_F)}{\mathcal{F}_{-1/2}(\eta_F)} \mu_n} \quad (25b)$$

In some prior publications, an incorrect version of eqns. (25) has been presented. In [6], eqn. (D10) is an incorrect version of (25). Reference [7] makes use of the incorrect equation in eqn. (A8) [7]. The correct relation between average MFP for backscattering and mobility is given by eqns. (25), which assumes a 2D semiconductor with parabolic energy bands and one occupied subband. It could, with added complexity, be generalized to treat nonparabolic energy bands and multiple occupied subbands.

## 5. Power Law Scattering

To proceed further, we need to specify the energy-dependent MFP. The simplest way to do this is to assume **power law scattering** and write ([4], eqn. (3.31))

$$\lambda(E) = \lambda_0 \left( \frac{E - E_C}{k_B T} \right)^r . \quad (26)$$

Using eqns. (21), (24) and (26) in (16a), we find an expression for the diffusive mobility in terms of the MFP prefactor, $\lambda_0$, and the MFP characteristic exponent, $r$. The result is

$$\mu_n = \frac{\lambda_0 \upsilon_T}{2(k_B T/q)} \left( \frac{\Gamma(r+3/2)}{\Gamma(3/2)} \right) \left( \frac{\mathcal{F}_{r-1/2}(\eta_F)}{\mathcal{F}_0(\eta_F)} \right), \quad (27)$$

where ([4], eqn. (3.69))

$$\upsilon_T = \sqrt{\frac{2 k_B T}{\pi m^*}} . \quad (28)$$

It is interesting to note that if $r = 1/2$, then the mobility is independent of the location of the Fermi level.

Similarly, using eqn. (16b), we find an expression for the ballistic mobility in terms of the channel length, $L$. The result is

$$\mu_B = \frac{L\upsilon_T}{2(k_B T/q)} \left( \frac{\mathcal{F}_{-1/2}(\eta_F)}{\mathcal{F}_0(\eta_F)} \right). \tag{29}$$

Assuming power law scattering and parabolic energy bands (i.e. eqns. (22) and (26)), we can evaluate the average mean-free-path for backscattering from eqn. (19) and find

$$\langle\langle \lambda(E) \rangle\rangle = \lambda_0 \times \left( \frac{\Gamma(r+3/2)}{\Gamma(3/2)} \right) \frac{\mathcal{F}_{r-1/2}(\eta_F)}{\mathcal{F}_{-1/2}(\eta_F)}. \tag{30}$$

When $r = 0$ (no energy dependence to the MFP), then $\langle\langle \lambda(E) \rangle\rangle = \lambda_0$.

Note that eqn. (30) should also be eqn. (6.39) in [4], but there is an error in (6.39). The $\mathcal{F}_{+1/2}(\eta_F)$ in the denominator of (6.39) should be replaced by $\mathcal{F}_{-1/2}(\eta_F)$. The correct expression is (30) above or (3.32) in [4].

## 6. Validity of Matthiessen's Rule for the Ballistic Mobility
It should be noted that Matthiessen's Rule ([4], eqn. (3.39)),

$$\frac{1}{\mu_{app}} = \frac{1}{\mu_n} + \frac{1}{\mu_B}, \tag{31}$$

**does not follow in general** from eqn. (17). Only when the MFP is independent of energy, (i.e. $r = 0$ so $\lambda(E) = \lambda_0$) does eqn. (31) follow from eqn. (17).

In the general case, the expression for the apparent MFP involves the MFP prefactor, $\lambda_0$, the MFP characteristic exponent, $r$, the channel length, $L$, as well as the location of the Fermi level. Unfortunately, we cannot analytically integrate eqn. (17), except for the simple case of $r = 0$ (an energy independent MFP). In that case, the apparent mobility is given by Matthiessen's Rule, eqn. (31). In practice, one often uses Matthiessen's Rule even when the MFP is energy dependent. It gives the correct answer for the ballistic and diffusive limits, but can be in error in between those two limits

How much error do we make when we assume Matthiessen's Rule? We can answer this question as follows. Assume a power law scattering with a typical MFP prefactor. Next we need to determine the characteristic exponent, $r$. The MFP is proportional to velocity times scattering time ([4], eqn. (6.16)). For parabolic energy bands, velocity goes as the square root of energy. The scattering rate (one over the scattering time) is proportional to the density of states, which is independent of energy in 2D, so we conclude that $r = 1/2$. Finally, we need to assume an $n_S$ so that $\eta_F$ can be determined from eqn. (24). Now we can compute the apparent mobility vs. channel length from eqn. (17).

To compare the exact answer to Matthiessen's rule, we can separately compute $\mu_n$ from eqn. (16a) and $\mu_B(L)$ from eqn. (16b) and combine them according to eqn. (31). Comparing the exact answer obtained by integrating eqn. (17) to Matthiessen's Rule provides an estimate of the errors involved. Figure 1 below shows some results. We assume Si parameters with $m^* = 0.19 m_0$, $n_S = 10^{13}$ cm$^{-2}$ and $\lambda_0 = 10$ nm. We assume three values of the characteristic exponent for scattering in eqn. (21), $r = 1/2, 1$, and $3/2$. Figure 1 shows that the error in Matthiessen's rule can be significant when the scattering has a strong energy dependence, but for $r = 1/2$, the expected value, the error is only a few percent.

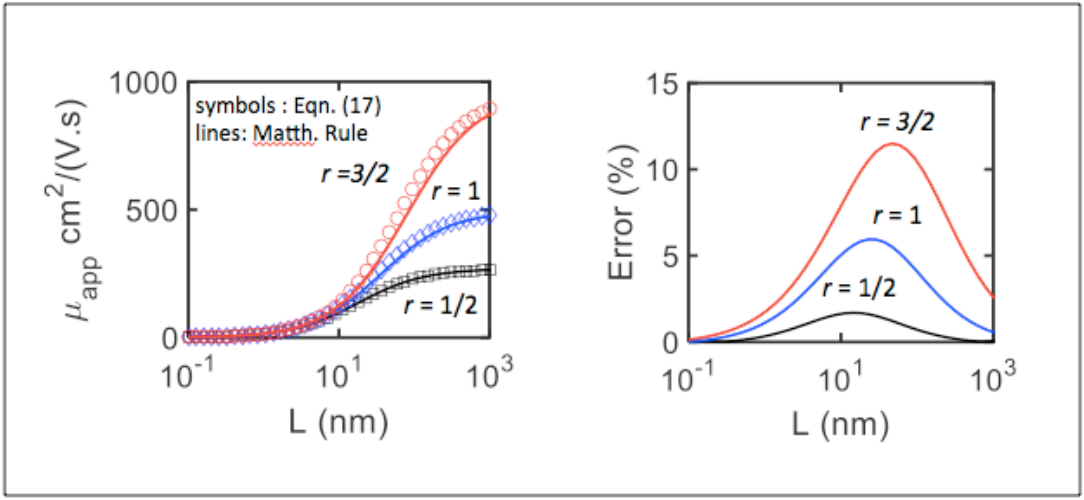

Fig. 1. Comparison of Matthiessen's rule for combining the ballistic and scattering limited mobility (eqn. (31)) and the exact answer, eqn. (17). Typical numbers for Si are assumed. Left: Apparent mobility vs. channel length. Right: Error in Matthiessen's rule vs. channel length.

## 7. Relation to the Drude Mobility
The Drude expression for mobility is

$$\mu_n = \frac{q \tau_0}{m^*}, \tag{32}$$

where $\tau_0$ is a constant relaxation time. Is this consistent with eqn. (27), which gives the mobility in terms of the mean-free-path and not the scattering time? In 2D, the energy-dependent mean-free-path for backscattering is ([4], eqn. (6.16),

$$\lambda(E) = \frac{\pi}{2} \upsilon(E) \tau(E). \qquad (33)$$

Assuming parabolic energy bands and a constant scattering time, we find

$$\lambda(E) = \frac{\pi}{2} \sqrt{2(E - E_C)/m^*} \, \tau_0, \qquad (34)$$

which can be written in power law form as in eqn. (26) with $r = 1/2$ and

$$\lambda_0 = \frac{\pi}{2} \sqrt{2k_B T/m^*} \, \tau_0. \qquad (35)$$

Using eqn. (35) in (27) and assuming $r = 1/2$ and nondegenerate carrier statistics, we find that eqn. (27) becomes eqn. (32). Our expression for mobility in terms of the MFP is equivalent to the Drude expression for mobility in terms of the scattering time – as it should be.

## 8. Summary

We have discussed several issues relevant to the transport of 2D carriers across the channel of a nanoscale field-effect transistor. In the process, we have pointed out an error in relating the average mean-free-path for backscattering to the mobility. This error occurred in in [6] and [7]; the correct expression is eqn. (25b). We also corrected an error in eqn. (6.39) of [4]; the correct result should is eqn. (30) here or eqn. (3.32) in [4].

For a 2D semiconductor with a single occupied subband, the results that are likely to be useful in analyzing experiments are listed below.

*Sheet carrier density:*

$$n_S = g_v \frac{m^* k_B T}{\pi \hbar^2} \mathcal{F}_0(\eta_F) \, \text{m}^{-2}, \qquad (24)$$

where

$$\eta_F = \frac{E_F - E_C}{k_B T} \qquad (23)$$

*Diffusive mobility and average mean-free-path for backscattering*

$$\mu_n = \frac{\upsilon_T \langle\langle \lambda(E) \rangle\rangle}{2(k_B T/q)} \frac{\mathcal{F}_{-1/2}(\eta)}{\mathcal{F}_0(\eta)} \qquad (25a)$$

*Diffusive mobility for power law scattering:*

$$\mu_n = \frac{\lambda_0 \upsilon_T}{2(k_B T/q)} \left( \frac{\Gamma(r+3/2)}{\Gamma(3/2)} \right) \left( \frac{\mathcal{F}_{r-1/2}(\eta_F)}{\mathcal{F}_0(\eta_F)} \right) \qquad (27)$$

*Ballistic mobility:*

$$\mu_B = \frac{L\upsilon_T}{2(k_B T/q)} \left( \frac{\mathcal{F}_{-1/2}(\eta_F)}{\mathcal{F}_0(\eta_F)} \right) \qquad (29)$$

*Matthiessen's Rule for ballistic and diffusive mobilities:*

$$\frac{1}{\mu_{app}} = \frac{1}{\mu_n} + \frac{1}{\mu_B} \qquad (31)$$

*Average MFP from mobility:*

$$\langle\langle \lambda(E) \rangle\rangle = \frac{(2k_B T/q)}{\upsilon_T} \frac{\mathcal{F}_0(\eta_F)}{\mathcal{F}_{-1/2}(\eta_F)} \mu_n \qquad (25b)$$